\documentclass[twoside,twocolumn,9pt]{article}
\usepackage{extsizes}
\usepackage[super,sort&compress,comma]{natbib}
\usepackage[version=3]{mhchem}
\usepackage[left=1.5cm, right=1.5cm, top=1.785cm, bottom=2.0cm]{geometry}
\usepackage{balance}
\usepackage{widetext}
\usepackage{times}
\usepackage{newtxtext,newtxmath}
\usepackage{eqnarray,amsmath}
\usepackage{sectsty}
\usepackage{graphicx}
\usepackage{booktabs}
\usepackage{lastpage}
\usepackage[format=plain,justification=raggedright,singlelinecheck=false,font={stretch=1.125,small,sf},labelfont=bf,labelsep=space]{caption}
\usepackage{float}
\usepackage{fancyhdr}
\usepackage{fnpos}
\usepackage[english]{babel}
\usepackage{array}
\usepackage{droidsans}
\usepackage{charter}
\usepackage[T1]{fontenc}
\usepackage[usenames,dvipsnames]{xcolor}
\usepackage{setspace}
\usepackage[compact]{titlesec}
\usepackage{bm}
\usepackage{multicol}

\usepackage{epstopdf}

\definecolor{cream}{RGB}{222,217,201}

\begin{document}



\makeFNbottom
\makeatletter

\def\red{\textcolor{red}}

\def\Gr{{\rm \Gamma}}
\newcommand{\be}{\begin{equation}}
\newcommand{\ee}{\end{equation}}
\newcommand{\bea}{\begin{eqnarray}}
\newcommand{\eea}{\end{eqnarray}}

\renewcommand\LARGE{\@setfontsize\LARGE{15pt}{17}}
\renewcommand\Large{\@setfontsize\Large{12pt}{14}}
\renewcommand\large{\@setfontsize\large{10pt}{12}}
\renewcommand\footnotesize{\@setfontsize\footnotesize{7pt}{10}}
\makeatother

\renewcommand{\thefootnote}{\fnsymbol{footnote}}
\renewcommand\footnoterule{\vspace*{1pt}%
\color{cream}\hrule width 3.5in height 0.4pt \color{black}\vspace*{5pt}}
\setcounter{secnumdepth}{5}

\makeatletter
\renewcommand\@biblabel[1]{#1}
\renewcommand\@makefntext[1]%
{\noindent\makebox[0pt][r]{\@thefnmark\,}#1}
\makeatother
\renewcommand{\figurename}{\small{Fig.}~}
\sectionfont{\sffamily\Large}
\subsectionfont{\normalsize}
\subsubsectionfont{\bf}
\setstretch{1.125} 
\setlength{\skip\footins}{0.8cm}
\setlength{\footnotesep}{0.25cm}
\setlength{\jot}{10pt}
\titlespacing*{\section}{0pt}{4pt}{4pt}
\titlespacing*{\subsection}{0pt}{15pt}{1pt}

\fancyfoot{}
\fancyfoot[LE]{\footnotesize{\sffamily{\hspace{2pt}\thepage}}}
\fancyfoot[RO]{\footnotesize{\sffamily{\hspace{2pt}\thepage}}}

\fancyhead{}
\renewcommand{\headrulewidth}{0pt}
\renewcommand{\footrulewidth}{0pt}
\setlength{\arrayrulewidth}{1pt}
\setlength{\columnsep}{6.5mm}
\setlength\bibsep{1pt}

\makeatletter
\newlength{\figrulesep}
\setlength{\figrulesep}{0.5\textfloatsep}

\newcommand{\topfigrule}{\vspace*{-1pt}%
\noindent{\color{cream}\rule[-\figrulesep]{\columnwidth}{1.5pt}} }

\newcommand{\botfigrule}{\vspace*{-2pt}%
\noindent{\color{cream}\rule[\figrulesep]{\columnwidth}{1.5pt}} }

\newcommand{\dblfigrule}{\vspace*{-1pt}%
\noindent{\color{cream}\rule[-\figrulesep]{\textwidth}{1.5pt}} }

\makeatother

\twocolumn[
  \begin{@twocolumnfalse}
\sffamily
\begin{tabular}{m{0.0cm} p{16.0cm} }

\quad &
\noindent\LARGE{\textbf{Modelling Bi-specific Antibodies in Aqueous Solution}} \\
\vspace{0.3cm} & \vspace{0.3cm} \\
& \noindent\large{Taras Hvozd$^1$, Yurij V. Kalyuzhnyi$^1$, Vojko Vlachy$^2$} \\
\quad & \textit{$^1$~Institute for Condensed Matter Physics of the National Academy of Sciences of Ukraine 1 Svientsitskii St., Lviv, Ukraine 79011;} \\ 
&\textit{E-mail: tarashvozd@icmp.lviv.ua,$\;\;$yukal@icmp.lviv.ua} \\ 
\quad & \textit{$^2$~Faculty of Chemistry and Chemical Technology, University of Ljubljana, Ve\v{c}na pot 113, SI--1000 Ljubljana, Slovenia. } \\ 
&\textit{E-mail: vojko.vlachy@fkkt.uni-lj.si} \\  
\vspace{0.3cm} & \vspace{0.3cm} \\
\quad 
& \noindent\normalsize
{This study presents theoretical results for physico-chemical properties of system of molecules modeling bi-specific antibodies, such as, dual-variable-domain monoclonal antibodies (DVD-Ig) and Fabs-In-Tandem Immunoglobulin (FIT-Ig). These molecules are representatives of the engineered proteins that combine the function and specificity of two monoclonal antibodies.  Individual  molecules are here depicted as an assembly of nine (or in case of the Fit-Ig eleven) hard spheres, organized  to resemble the Y-shaped object. The effects of the increased size, asymmetry, and flexibility of individual molecules on measurable properties of such systems of molecules are investigated. We examined the liquid-liquid phase separation, the second virial coefficient $B_2$, and viscosity under various experimental conditions. The calculations are compared with the data for regular monoclonal antibodies and discussed in view of the experimental results for DVD-Ig solutions available in literature.}
\end{tabular}

 \end{@twocolumnfalse} \vspace{0.6cm}

]

\renewcommand*\rmdefault{bch}\normalfont\upshape
\rmfamily
\section*{}
\vspace{-1cm}


\section{Introduction}

\noindent

Monoclonal antibodies (mAbs) are proteins of the immune system, engineered to suit therapeutic purposes ~\cite{Thayer2016,Reichert2017,Starr2019}. They offer promising treatments for auto-immune disorders, cancer, and many other often fatal diseases. In developing of therapeutic proteins and their mixtures~\cite{Krieg2020} the goal is to obtain formulations that are sufficiently concentrated in antibodies and yet free of protein aggregates, see 
Ref.~\cite{Basle2020}. The latter, not only limit their effectiveness as therapeutics, but may also pose a health risk for patients.  

Single-target immunotherapy does not seem to destroy sick cells sufficiently. Binding multiple epitopes with a single (bi-specific) antibody offers significant benefits, the most important one is much higher specificity. For review on possible future developments and application of these proteins see references~\cite{Kontermann2015,Ma2021,Gong2017,Gong2019,Svilenova2023}. 
In simple words we may say that bi-specific antibodies (bsAbs) combine the functionality of two antibodies in one molecule.

In literature we can find several Ig molecular platforms~\cite{Ma2021}, one is DVD-Ig and the other is Fit-Ig.  
This new generation of therapeutics can target two or even more disease mechanisms.  They are designed for cancer immunotherapy~\cite{Chames2009} and currently several of them are in clinical use as also in diagnostics. They are designed as an Ig-like molecules, except that each light chain and heavy chain contains two variable domains. As such they are larger than an original immunoglobulins, asymmetric and in addition also more flexible~\cite{Correia2013}. Unfortunately, this increases their propensity for aggregation~\cite{Starr2019,Svilenova2023}. 

In the present study we propose to analyse some measurable properties of the model DVD-Ig (dual variable domain immunoglobulin)
~\cite{DiGiammarino2012,Jakob2013,Correia2013,Raut2015,Raut2016,Raut2016a} and FIT-Ig proteins in aqueous environment. These large antibodies (M$_W$ around 200 kDa and more), having increased asymmetry and flexibility, should be costly to simulate in detail. Actually, we do not know any such attempt. To examine the effects of the increased size and flexibility on thermodynamic properties we utilized here an  
improved version of the thermodynamic perturbation theory (TPT)
for associating fluids~\cite{Butovych2023}. 

Original version of the theory~\cite{Wertheim3,Wertheim4,Wertheim1987} has been successfully applied in several recent studies~\cite{Kastelic2016b,Kastelic2017,Kastelic2018,Hvozd2020}.
The theoretical improvements presented here, enables us to account for the difference in the bonding abilities of the square-well
sites located on the surface of different hard-sphere monomers of the molecule. This difference appears as a result of blocking effects due to the rest of the hard-sphere monomers forming the molecule~\cite{Butovych2023}.

As main results of our calculations we present the viscosity data for model proteins,
the second osmotic virial coefficient and the liquid-liquid phase separation diagrams. We discuss the results in view of experimental data available in literature~\cite{Raut2015} and compare our findings with the results for regular mAbs studied recently~\cite{Kastelic2016b,Kastelic2017,Kastelic2018,Hvozd2020}.

\section{Models of the bispecific antibodies solution~\label{sec:Model}}

\begin{figure}[htb!]
	\centering
	\includegraphics[width=3.4in]{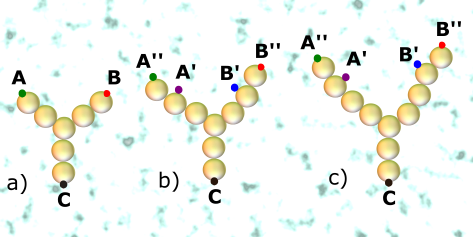}
	\caption{\small Models of the antibodies studied here are: (a) Regular 7-bead monoclonal antibody ~\cite{Kastelic2017}, (b) DVD-Ig antibody (9-bead model), and (c) FIT-Ig antibody (11-bead model) \cite{Ma2021}. For an additional illustration see also Figs. 1 and 1A of Refs.~\cite{DiGiammarino2012,Correia2013}.} 
	\label{Figure1}
\end{figure}

The 9- and 11-bead representations of the antibodies studied here are modifications of the 7-bead model proposed and applied before~\cite{Kastelic2017,Kastelic2018,Hvozd2020}: the models are schematically presented in Figure~\ref{Figure1}.
Such asymmetric (Y-like shape) "proteins" are constructed from nine or eleven (not seven as are regular mAbs) beads. For additional illustrations see also Fig. 1 of Ref.~\cite{Ma2021} and Fig. 1 of Ref.~\cite{Correia2013}.

We consider a solution of protein molecules, schematically shown in Figure \ref{Figure1}, 
with the number density $\rho_s$. Molecules modeling antibodies are depicted by 
{$n_s$} tangentially bonded hard-sphere beads (see also Ref. \cite{Kastelic2018}), forming three-arm completely flexible  star shaped molecules with $l_s$-bead arms attached to the central sphere. There is $l_s$=2  beads forming the base and $l_s$=3 and $l_s=4$ beads for the two upper arms  of DVD-Ig and FIT-Ig antibody models, respectively (Figure \ref{Figure1}). 
Hard-sphere diameters of the beads composing the model protein are denoted $\sigma_s$.
{
	The last two monomers of the three- and four-bead arms, which represent double variable domains, each bear one off-center square-well site, randomly placed on the surface. These sites 
	are denoted as $A"$ and $A'$ (for double variable Fab domain) and $B"$ and $B'$ 
	(for double variable Fab' domain). As before, there is only one off-center square-well
	site located on the terminal bead of the arm representing fragment crystallizable region
	and denoted as $C$. 
	The pair potential $U_{KL}(12)$ between these beads is
	\be
	U_{KL}(12)=U_{hs}(r)+U_{KL}^{(as)}(12),
	\label{Uas}
	\ee
	where
	\be
	U^{(as)}_{KL}(12)=\left\{
	\begin{array}{rl}
		-\epsilon_{KL}, & {\rm for}\;z(12)\le\omega\\
		0, & {\rm otherwise}
	\end{array}
	\right.,
	\label{UassH} 
	\ee 
	$U_{hs}(r)$ is the hard-sphere potential, $K,L=A",A',B",B',C$,  $\omega$ and 
	$\epsilon_{KL}(>0)$ are the width and depth of the off-center square-well sites, respectively,
	and $z(12)$ is the distance between these square-well sites. 
}

Effects of the size, asymmetry and flexibility of such molecules on the physico-chemical properties of their solutions are investigated next. More precisely, we studied the liquid-liquid phase separation, the second virial coefficient, $B_2$, and viscosity, all under various experimental conditions. The results are compared with the data obtained for regular (7-bead) monoclonal antibodies and discussed in view of the experimental results for DVD-Ig solutions available in literature~\cite{Raut2015,Raut2016,Raut2016a}. 

\section{Calculation of measurable quantities}

{
	Theoretical approach used here~\cite{Wertheim3,Wertheim4,Wertheim1987} has previously been explained in several papers, see, for example,~\cite{Kastelic2017,Kastelic2018,Hvozd2020,Hvozd2022}.
	Similarly as before, the Helmholtz free energy of the system $A$ is presented as a sum of two terms,
	\be
	A=A_{ref}+\Delta A_{as}
	\label{free}
	\ee
	where $A_{ref}$ is 
	the Helmholtz free energy of the reference system and $\Delta A_{as}$ the corresponding
	contribution due to protein association. Here the reference system is represented by the $n_s$-bead
	model with $\epsilon_{KL}=0$. According to our previous studies 
	~\cite{Kastelic2017,Kastelic2018,Hvozd2020,Hvozd2022} we have
	\be
	{\beta A_{ref}\over V}=\rho_s[\ln(\rho_s\Lambda^3)-1]-(n_s-1)\rho_s\ln g^{(PY)}_{hs}
	+{\beta \Delta A_{hs}\over V},
	\label{Aref}
	\ee
	where $g^{(PY)}_{hs}=g^{(PY)}_{hs}(\sigma^+_s)$ is the Percus-Yevick contact value of the hard sphere radial distribution function (RDF) and
	$\Delta A_{hs}$ is excess hard-sphere Helmholtz free energy. $\Delta A_{as}$ is calculated using expression similar to that used earlier i.e.
	\be
	{\beta \Delta A_{as}\over V}=\rho_s\left(\sum_{K=A_1}^C\ln{X_K}-
	{1\over 2}\sum_{K=A_1}^C X_K+{5\over 2}\right),
	\label{Aas}
	\ee
	where $K$ takes the values $A_1,A_2,B_1,B_2,C$ and $X_K$ is fraction of the particles $not$
	bonded on a site $K$. These fractions follow from the solution of the set of five equations
	\be
	\rho_s X_K\sum_{L=A_1}^C \Delta_{KL}X_L+X_K-1=0,
	\label{XK}
	\ee
	where 
	\be
	\Delta_{KL}=4\pi g_{KL}^{(hs)}\int_{\sigma}^{\sigma_s+\omega}{\bar f}_{KL}(r)r^2dr,
	\label{D}
	\ee
	\be
	{\bar f}_{KL}(r)=\left(e^{\beta\epsilon_{KL}}-1\right)\left(\omega+\sigma_s-r\right)^2
	\left(2\omega-\sigma_s+r\right)/\left(6\sigma_s^2r\right),
	\label{fbar}
	\ee
	and $g_{KL}^{(hs)}=g_{KL}^{(hs)}(\sigma_s^+)$ is the contact value of site-site RDF between the  centers of the hard-sphere beads bearing sticky sites of type $K$ and $L$ \cite{Kalyuzhnyi1995,Kalyuzhnyi1997,Lin1998,Butovych2023}, i.e.
	\be
	g_{KL}^{(hs)}=g_{hs}+{1\over 2(\phi-1)},\;\;(K=A",B",C;\;L=A",B",C)
	\label{gAB}
	\ee
	\be
	g_{KL}^{(hs)}=g_{LK}^{(hs)}=g_{hs}+{3\over 4(\phi-1)},\;\;
	(K=A",B",C;\;L=A',B')
	\label{gAA}
	\ee
	\be
	g_{KL}^{(hs)}=g_{hs}+{1\over \phi-1},\;\;(K=A',B';\;L=A',B')
	\label{gBB}
	\ee
	where $\phi=n_s\pi\sigma_s^3\rho_s/6$. Note that in the framework of the regular TPT for associating fluids, expression for the contact value of hard-sphere RDF $g_{hs}$ 
	in the corresponding expression for $\Delta_{KL}$ is 
	used~\cite{Wertheim3,Wertheim4,Wertheim1987,Kastelic2017,Kastelic2018,Hvozd2020,Hvozd2022}.
	This modification of the TPT enable us to account for the blocking effects due to the nearest neighboring monomers forming a chain \cite{Butovych2023}.
	
	From here on, the chemical potential $\mu$ and pressure $P$ can be obtained via the standard relations, i.e.
	\be
	\mu=\left({\partial(A/V)\over\partial\rho_s}\right)_{T,V},\;\;\;P=\rho_s\mu-A/V
	\label{muP}
	\ee
	and used for the phase equilibrium calculations. The densities of the coexisting phases follow from the solution of the set of two equations
	\be
	\mu(T,\rho_s^{(g)})=\mu(T,\rho_s^{(l)})
	\label{mueq}
	\ee
	\be
	P(T,\rho_s^{(g)})=P(T,\rho_s^{(l)}),
	\label{Peq}
	\ee
	representing phase equilibrium conditions. Here, $\rho_s^{(g)}$ and $\rho_s^{(l)}$ are the densities of low-density and high density phases, respectively.}

Our calculation of the viscosity is carried out following the scheme developed earlier. According to this scheme \cite{Kastelic2017} the relative viscosity of a solution, $\eta/\eta_0$, can be given as
\be
\ln{\left({\eta\over\eta_0}\right)}=\sum_{n=1}^\infty\gamma f(n) P(n,\gamma),
\label{relative}
\ee
where $\eta$ is viscosity of the solution and $\eta_0$ viscosity of the solvent, while $\gamma$ is the mass concentration of solute molecules (mg of protein per mL of the solution). Further, $P(n,\gamma)$ is the weight fraction 
distribution of the clusters containing $n$ molecules, while $f(n)$ describes contribution of the cluster of the size $n$ to the viscosity. For $f(n)$ we use here the function applied previously \cite{Kastelic2017},
i.e.
\be
f(n)=c n^d,
\label{f}
\ee
where $c$ and $d$ are the fitting parameters, obtained from the comparison of the theoretical and experimental data.
We assume that the sites $A",B"$ and $C$, which are located on the tips the variable and crystallizable domains, interact in-between with the same strength. Also, the sites $A',B'$, which are located on the monomers next to the terminal monomers 
of the variable domains, interact in-between and with the rest of the sites with reduced strength, i.e. $\epsilon_{KL}=\epsilon$ for $K,L=A",B",C$, $\epsilon_{KL}=\epsilon_{LK}=k\epsilon$ for $K=A',B'$ and $L=A",B",A',B',C$. Notice that $0\leq k \leq 1$.

The distribution function $P(n,\gamma)$ can be calculated using the information about fractions of the molecules $X_K$ not boded through the site $K$. For the model at hand we have a set of five fractions, which completely define $P(n,\gamma)$, i.e. $X_{A'}$, $X_{B'}$, $X_{A''}$, $X_{B''}$ and $X_C$. Note that due to the symmetry of the model we have: $X_{A"}=X_{B"}=X_{C}$ and $X_{A''}=X_{B''}$. Next we assume that distribution function $P(n,\gamma)$ can be approximated by the distribution function of the version of the model with three effective equivalent sites, i.e.
\be
P(n,\gamma)={\tilde P}(n,\gamma),
\label{P1}
\ee
where \cite{Bianchi2006}
\be
{\tilde P}(n,\gamma)={3(2n)!\over (n+2)!(n-1)!}X^3\left[(1-X)X\right]^{n-1}
\label{P2}
\ee
and $X$ is the fraction of the particles, which are not bonded via one of these sites.
To relate both versions of the model we assume that this fraction is obtained from the equality
of Helmholtz free energy for the current model and the model with three equivalent sites, i.e.
\be
\Delta{\tilde A}_{as}=\Delta A_{as},
\label{XAas}
\ee
where for $\Delta A_{as}$ we  use the expression (\ref{Aas}) and 
\be
{\Delta{\tilde A}_{as}\over V}=\rho_s\left(3\ln{X}-{3\over 2} X+{3\over 2}\right).
\label{Xeq}
\ee
Due to this relation thermodynamics of the model with equivalent sites is the same as that of the original model. 
Note that for $k$=0 the five-site model reduces to the model with three equivalent sites. 


\bigskip
\section{Results and comparison with experimental data}

\begin{table}[h]

	\caption{Parameters of the models for mAb and DVD-Ig protein solutions}\
	\begin{tabular}{|c|c|c|c|c|c|}
		\hline
		protein & $p$H & $\;I\;$ & $\epsilon$ & $\;c\;$ & $\;\;d\;\;$ \\
		&  & $mM$& $kJ/mol$ & $\;mL/mg\;$ & $\;\;\;\;$ \\
		\hline
		& & & & & \\ 
		mAb  & 6.1      & 0.0  &   33.26   & 0.0220 & 0.44\\
		-  & 6.1      & 15.0   &  23.28   & 0.0220 & 0.44 \\
		& & & & & \\
		DVD-Ig  & 5.1      & 15.0 &   33.26   & 0.0255 & 0.44\\
		-  & 5.1      & 50.0   &   30.76   & 0.0255 & 0.44\\
		& & & & & \\
		-  & 6.1      & 0.0   &   42.07   & 0.0255 & 0.44 \\
		-  & 6.1      & 15.0  &   40.99   & 0.0255 & 0.44 \\
		-  & 6.1      & 50.0  &   39.66   & 0.0255 & 0.44 \\
		& & & & & \\
		-  & 6.5      & 15.0  &   41.41   & 0.0255 & 0.44\\
		-  & 6.5      & 50.0  &   39.41   & 0.0255 & 0.44\\
		& & & & & \\
		-  & 7.0      & 15.0  &   40.91   & 0.0255 & 0.44\\
		-  & 7.0      & 50.0  &   38.66   & 0.0255 & 0.44\\
		\hline
	\end{tabular}
	\label{table1}
\end{table}
In important experimental papers Raut and Kalonia~\cite{Raut2015,Raut2016,Raut2016a} examined the liquid-liquid phase separation in solutions of DVD-antibodies as also their viscosities. 
They measured the cloud-point temperature, $T_{cloud}$, which is marking the onset of the phase separation. They have investigated the effects of size of the protein, the salt concentration, $p\mathrm{H}$, and nature of some other additives {on the phase behavior and viscosity of the solution}. Unfortunately, our present model does not include electrostatics so it cannot account for these effects.  We can therefor only investigate the influence of the increased size and asymmetry of the DVD molecule in comparison with regular antibody molecule studied before~\cite{Kastelic2017,Kastelic2018}.

\subsection{Viscosity}

We fit the model calculations to experimental data for the viscosity of antibody solutions presented in \cite{Raut2016}.
In that paper viscosity measurements for mAb and DVD-Ig protein 
solutions at different values of $p$H, ionic strength $I$ and protein mass concentration $\gamma=M_2\rho_s/N_A$ were carried out. the molecular weights of mAb and DVD protein molecules were $M_2=150,000$ g/mol and $M_2=200,000$ g/mol, respectively, and $N_A$ denoted Avogadro's number. The temperature of the solutions was $25^\circ$C.

The model parameters, which have to be fitted are: $\omega$, $\sigma$, $c$, $d$, $\epsilon$ and $k$.  The value of $\omega$ was as before chosen to
be approximately equal to the hydrogen bond length, i.e. $\omega=0.15$ nm. Further we assume that the size of hard-sphere monomers $\sigma$ for both 7-bead and 9-bead models is the same and we are using here the value $\sigma=2.5$ nm. Also, in all the cases studied here we use the same value for the parameter $k$ defining the strength of the cross interaction; $k$=0.6. 
Parameters $c$ and $d$ describe contribution of clustering to viscosity of the solution. 
In this analysis we assume that these parameters depend only on the type of
the molecules (either mAb or DVD-Ig) and are independent of the ionic strength of the
solution and its $p$H value. This is a severe approximation.

\begin{figure}[ht] 	
	\begin{center}
		\includegraphics[width = 8.5 cm]{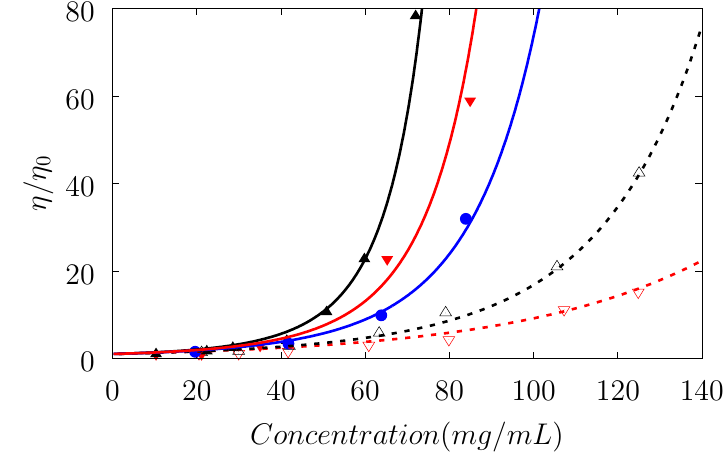}
		\caption{Relative viscosity $\eta/\eta_0$ of mAb (dashed lines and empty triangles) 
			and DVD-Ig (solid lines and filled triangles and circles) protein solutions as a 
			function of protein mass concentration $\gamma$ at ionic strength $I=0$ (upward triangles
			and black lines), $I$=15 mM (downward triangles and red lines) and $I$=50 mM (circles and
			blue line) at $T=25^\circ$C and $p$H=6.1. Symbols represent experimental 
			\cite{Raut2016a} and and lines the theoretical (this work) results.}
		\label{f_visc11}
	\end{center}
\end{figure}

\begin{figure}[ht] 	
	\begin{center}
		\includegraphics[width = 8.5cm]{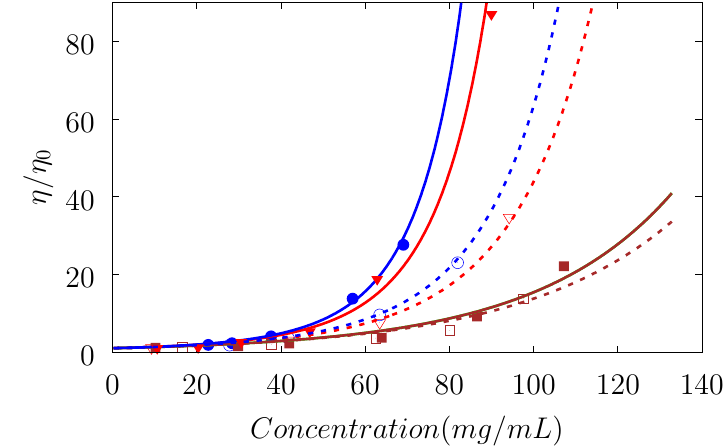}
		\caption{Relative viscosity $\eta/\eta_0$ of DVD-Ig protein solution $vs$ 
			protein mass concentration $\gamma$ at ionic strength $I$=15 mM (solid lines and filled symbols) and $I$=50 mM (dashed lines and open symbols) for $p$H=5,1 (brown lines and squares), $p$H=6.5 (blue lines and circles) and $p$H=7.0 (red lines and triangles). Symbols represent experimental \cite{Raut2016a} and lines our theoretical results.}
		\label{f_visc31}
	\end{center}
\end{figure}

The choice of parameters used in calculations is collected in Table \ref{table1} and 
comparison of the experimental and theoretical results for viscosity 
are presented in figures \ref{f_visc11}-\ref{f_visc31}. We note in passing that the chosen set of parameters is not unique. 
We have carefully examined the effects of reasonable variations of these parameters; small quantitative but not qualitative changes were noticed. 


In Figure \ref{f_visc31} we show our results and corresponding experimental data\cite{Raut2016a} for DVD-Ig solution at different values of $p$H
and ionic strength $I$, i.e. $p$H=5.1, 6.5, 7.0 and $I$=15 mM, 50 mM. 
The conclusion is that the model can describe viscosity measurements reasonably well. Additional information comes from inspection of Table 1. 
Though our model does not account for electrostatics we can still learn some useful information from the variation of parameters in Table 1. 
It is interesting that $c$ and $d$ do not change with $p$H of the solution neither with the  added  salt concentration so the results in figures \ref{f_visc11} and \ref{f_visc31} are fully determined by the parameter $\epsilon$, i.e. by the protein-protein attraction. 
In other words, viscosity of the DVD-Ig solutions measured in Ref. \cite{Raut2016a} can be modelled with reasonable accuracy fixing values of all other parameters but the energy of patch-patch attraction, $\epsilon$. Note that this parameter is defined as negative quantity (see Eq. \ref{UassH}) so smaller in magnitude value of $\epsilon$ means weaker attraction between the protein sites (patches). For all the situations analyzed here the attraction decreases with an increasing salt content. This holds true for mAbs and DVD proteins. 

\subsection{Second virial coefficient}

The second virial coefficient, $B_2$, quantifying the binary solute-solute interaction in dilute solutions, is one of the most important measurable quantities in protein solutions. It is known that value of this parameter can be used as an indicator of the crystallization \cite{George1994} as also, that low $B_2$ values are indicative for high viscosity~\cite{Tomar2018,Kastelic2017}. This coefficient is defined as
\be
\frac{\beta \Pi}{\rho_s}  =  1 + B_2\rho_s + O(\rho_s^2),
\label{b2}
\ee
and can be obtained from the osmotic pressure equation as explained elsewhere~\cite{Kalyuzhnyi2016}. It is most often presented as a function of the protein concentration {$\rho_s$.}

\begin{figure}[h] 	
	\begin{center}
		\includegraphics[width = 8.8 cm]{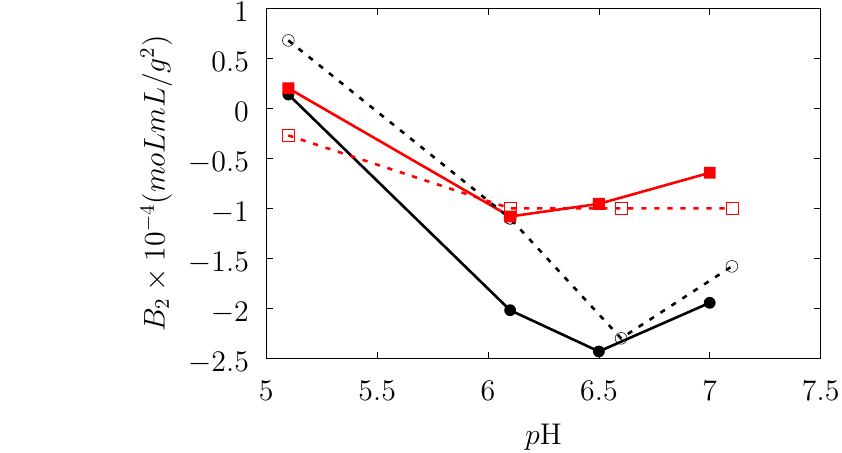}
		\caption{Second virial coefficient $B_2$ vs $p$H for the DVD-Ig solution at ionic strength $I=15$ mM (black 
			lines and symbols) and $I$=50 mM (red lines and symbols). Here solid lines and filled symbols represent 
			theoretical results and dashed lines and open symbols stand for the experimental results \cite{Raut2016a}.}
		\label{B22}
	\end{center}
\end{figure}

In contrast to this here in Figure \ref{B22} we compare our predictions for the second virial coefficient 
of the DVD-Ig solution as a function of $p$H, but at two different values of the ionic strength, i.e. $I$=15 and $I$=50 \cite{Raut2016a}. We found a qualitative agreement
between our calculations and experimental data. 
At $p$H=5.1 experimental measurements give
small positive and small negative values for $B_2$ at $I=15$ mM and $I=50$ mM,
respectively. At the same time for this value of $p$H and both values of the
solution ionic strength theory predict almost the same small positive values,
which are intermediate between those obtained experimentally. As $p$H 
increases, both experimentally and theoretically calculated $B_2$ decrease until
around $p$H$\approx$6.1 for $I=50$ mM and $p$H$\approx$6.5 for $I=15$ mM, where
$B_2$ is negative and reach its minimum values. Thus at these values of $p$H
interaction between the protein molecules is strongly attractive. We need to know that p$I$ of the protein is around 7.5.

This is also reflected in the behavior of the viscosity as a function of $p$H,
shown in Figures \ref{f_visc11} and \ref{f_visc31}. Here the most rapid increase of
the viscosity is observed for $p$H=6.5. On the other hand while experiment predict 
almost the same values of $B_2$ for solutions with $I=15$ and $I=50$ at $p$H=6.1
(see Figure \ref{B22}) corresponding values of the viscosity under the same conditions 
are quite different (see Figure \ref{f_visc11} and Figure 6 of \cite{Raut2016a}). Here 
viscosity of the solution with $I$=15 is about two times larger than that with $I$=50.
This is reflected in the behavior of the theoretically calculated $B_2$, i.e. here
$B_2(I=15)\approx 2B_2(I=50)$. Further increase of $p$H causes
slight increase of the theoretically calculated $B_2$, which is still has a negative value. Similar behavior can be observed for the experimentally obtained $B_2$ for
the solution with $I$=15 mM. For solution with $I$=50 mM experimental $B_2$ remains
constant for $p$H values in the range $6,1\leq p{\rm H}\leq 7.1$
Thus the model used here enables us to reproduce in general correlation between 
viscosity and second virial coefficient of the system. For more accurate description 
of this correlation less coarse grained model will be needed.

\subsection{Liquid--liquid phase separation}

Here we investigate the effects of the size (molecular mass) and asymmetry on the liquid--liquid phase separation. We examined three different models presented in Figure 1(a), (b), and (c); i.e. the regular mAbs, and two bi-specific variants of antibodies (b) DVD-Ig as also and (c) FIT-Ig. 

\begin{figure}[h] 	
	\begin{center}
		\includegraphics[width = 8. cm]{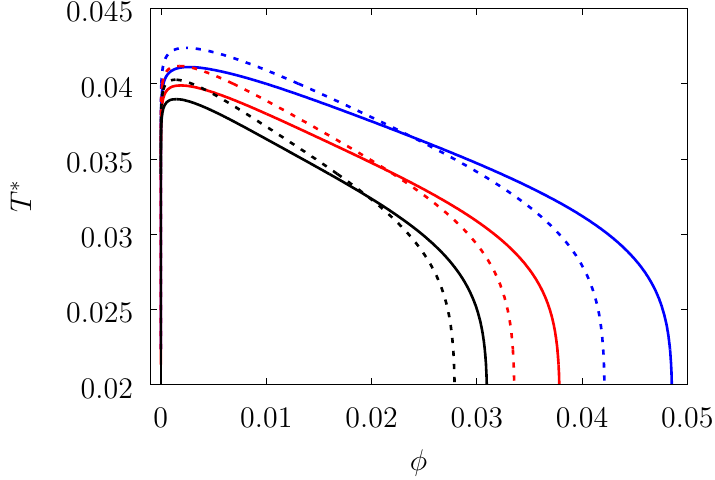}
		\caption{Liquid-liquid phase diagram in 
			$T^*=k_BT/\epsilon\;vs\;\phi=n_s\pi\rho_s\sigma^3/6$ coordinate frame
			for 7- 9-, and 11-bead models with $\epsilon_{KL}=\epsilon$
			for $K,L=A",B",C$ and $\epsilon_{KL}=\epsilon_{KC}=\epsilon_{CK}=0$
			for $K,L=A',B'$, calculated using TPT1 
			(dashed lines) and mTPT (solid lines). Here from the right to the 
			left at $T^*=0.025$: $n_s=7$ (blue lines), $n_s=9$ (red lines), $n_s=11$ 
			(black lines).}
		\label{mon_7_9_11-1}
	\end{center}
\end{figure}

\begin{figure}[h] 	
	\begin{center}
		\includegraphics[width = 8. cm]{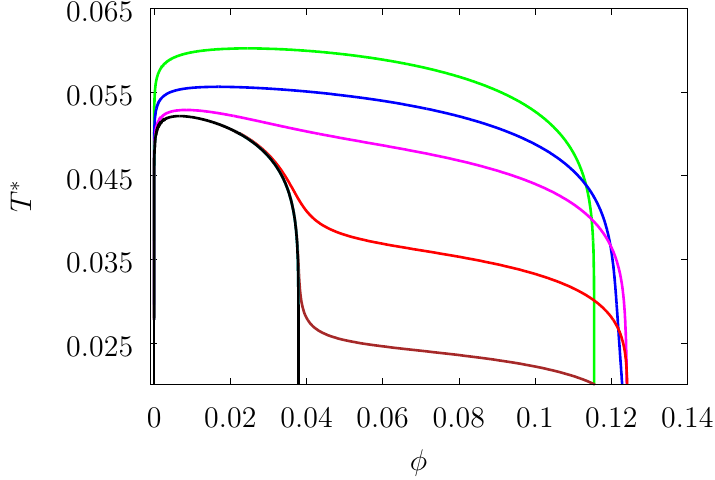}
		\caption{Liquid-liquid phase diagram for 9-bead model (DVD-Ig) with 
			$\epsilon_{KL}=\epsilon\;\;(K,L=A",B",C)$ and
			$\epsilon_{A'L}=\epsilon_{B'L}=k\epsilon\;\; (L=A",B",A',B',C)$
			in $T^*\;vs\;\phi$ coordinate frame. Here from the top to the bottom at $\phi=0.06$: $k$=1 (green line), $k=$0.9 (blue line), $k$=0.8 (pink line), $k$=0.6 (red line) and $k=$0.4 (brown line). Black line represent  results for $k$=0.}
		
		\label{phd_A1B1_0-1}
	\end{center}
\end{figure}

The calculations (see Figures \ref{mon_7_9_11-1} and \ref{phd_A1B1_0-1}) present data for the liquid--liquid phase separation modeling molecules shown in Figure 1. In this calculation all the proteins are interacting via the outermost beads: in case of (a) $\epsilon_{KL}=\epsilon$
($K,L=A,B,C$). For DVD-Ig and FIT-Ig model molecules (models (c) and (d)) we have:
$\epsilon_{KL}=\epsilon$ ($K,L=A",B",C$) and $\epsilon_{KL}=\epsilon_{KC}=\epsilon_{CK}=0$ 
($K,L=A',B'$). Numerical results for this case  are presented in 
Figure \ref{mon_7_9_11-1}. In this figure we show our results using regular TPT1 approach
\cite{Kastelic2017} as also its improved version, the so-called modified TPT (mTPT) \cite{Butovych2023}, as discussed above. 
Note that within the mTPT approach, instead of the contact values of the hard-sphere RDFs, the contact values of site-site RDFs (\ref{gAB})-(\ref{gBB}) were used. Here mTPT gives slightly 
wider phase diagrams with slightly smaller values of the critical temperatures.

In Figure \ref{phd_A1B1_0-1} we present our results for the phase diagram of 9-bead DVD-Ig model with different values of the patch-patch interaction (measured by parameter $k$), which include the patches of the type $A'$ and $B'$, i.e.
\be
\epsilon_{KL}=\epsilon\;\;(K,L=A",B",C)
\label{epsKL0}
\ee
and
\be
\epsilon_{A'L}=\epsilon_{B'L}=k\epsilon\;\; (L=A",B",A',B',C).
\label{epsL0}
\ee

\begin{figure}[h] 	
	\begin{center}
		\includegraphics[width = 8. cm]{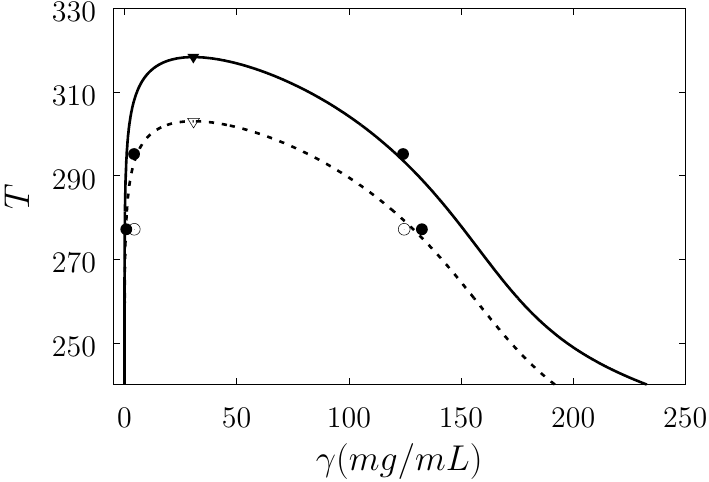}
		\caption{Liquid-liquid phase diagram for DVD-Ig protein solution in $T$ $vs$ 
			$\gamma$ coordinate frame at $p$H=6.5 and ionic strength $I$=15 mM (solid curve and filled  circles) and $I$=50 mM (dashed curve and empty circles). Here experimental results \cite{Raut2015} are shown by the circles and theoretical results are depicted by the lines. Theoretical predictions for the critical temperature and density are shown by empty ($I$=50 mM) and filled ($I$=15 mM) triangles, respectively.}
		
		\label{f_ph1_exp}
	\end{center}
\end{figure}
\begin{figure}[h] 	
	\begin{center}
		\includegraphics[width = 8. cm]{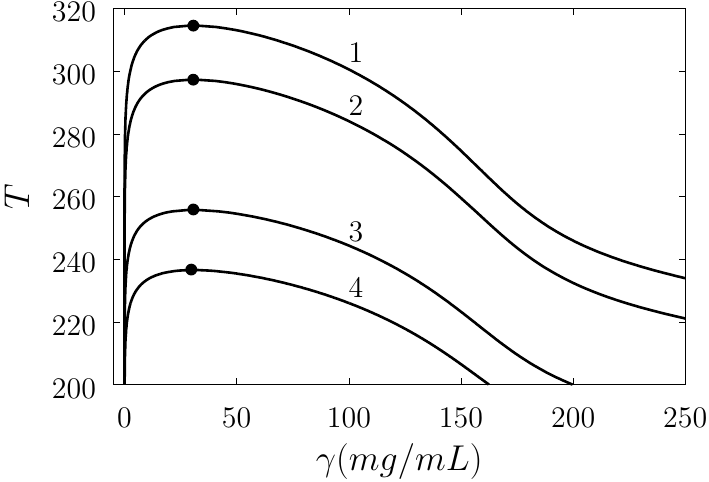}
		\caption{Liquid-liquid phase diagram for DVD-Ig protein solution in $T$ $vs$ $\gamma$ coordinate frame at $p$H=7.0 and ionic strength $I=15$ mM (1) 
			and $I=50$ mM (2) and at $p$H=5.1 and ionic strength $I=15$ mM (3) and $I=50$ mM (4). Here filled circles denote critical $(T,\gamma)$ points.}
		\label{f_ph2_exp}
	\end{center}
\end{figure}
Thus for $k$=1 we have the 5-patch model and for $k=0$ we have the 3-patch model.
For the parameter $k$ being smaller then one the sites are not completely coupled and the phase diagram gradually changes its shape. It becomes narrower around the critical temperature and wider for low temperatures, as it is shown in Figure \ref{phd_A1B1_0-1}. 
For$k\approx 0.6$ or less one can identify two distinct coexistence regions:
at higher temperatures the phase diagram is narrow and coincide with the phase
diagram for the three-patch version of the model and at lower temperatures the coexistence
region is about three times wider and coincide with the coexistence region of the five-patch version of the model. Transition between these two regions occurs in a narrow window
of the temperature with the width, which becomes smaller for smaller values of 
parameter $k$. 

Figures \ref{f_ph1_exp} and \ref{f_ph2_exp} present our predictions for the liquid-liquid phase behavior of the DVD-Ig protein solution at different values of $p$H and ionic strength.
The calculations were carried out for the same values of model parameters $\sigma$, $k$ and $\epsilon$ as determined in our viscosity calculations above. The values for the potential well depth $\epsilon$ are given in Table 1. 
However, the value of the potential well width $\omega$ needed to reproduce the experimental 
results for the liquid-liquid phase diagrams of the solution at $p$H=6.5 and two values of
the ionic strength, i.e. $I$=15 mM and $I$=50 mM \cite{Raut2015}, is in this case twice larger, i.e. $\omega$=0.3 nm (see Figure \ref{f_ph1_exp}). 
Unfortunately, the authors \cite{Raut2015} did not provide the liquid-liquid phase separation graph in the form T* vs protein concentration be compared with calculations in the broader range. Their experimental data~\cite{Raut2015} are presented in Figure 4.
Experiment predicted that there is no phase separation 
above $T=295^\circ$K for the solution with $I$=50 mM and above $T=298^\circ$K
for the solution with $I$=15 mM, so we may take this value as an estimate of the critical temperature. They also estimated for the critical density to be somewhere between 20 and 100 mg/ml. Theoretically calculated phase diagrams appear to be in a 
reasonable agreement with these values of the critical temperature and density.
For the solutions with $I$=50 mM and $I$=15 mM critical temperature is
$T_{cr}=303.0^\circ$K and $T_{cr}=318.4^\circ$K, respectively. In both cases the
critical densities are the same, i.e. $\gamma_{cr}=30.8$ mg/mL.

Similar values of the parameters were used to predict the phase behavior of DVD-Ig solutions at other conditions: $p$H=7.0 ($I$=15 mM and $I$=50 mM) and $p$H=5.1 ($I$=15 mM and $I$=50 mM). These results are shown in Figure \ref{f_ph2_exp}.
Here location of the critical temperature for the phase envelopes at different
values of $p$H and $I$ follows the general trend observed for the second virial
coefficient (see Figure \ref{B22}), i.e. $T_{cr}^{(1)}>T_{cr}^{(2)}>T_{cr}^{(3)}>T_{cr}^{(4)}$.

It has been shown by Bianchi and coworkers \cite{Bianchi2006} that for patchy colloids the width of the LLPS envelope critically depends on the number of patches (i.e. the number of the off-center square-well sites) on the particles. In our one-component model the minimum number of sites fully bonded is three for the system to be able to form a network and to phase separate. However, by studying binary mixtures with varying the average number of attractive square-well sites\cite{Bianchi2006} 
it has been possible to reduce the LLPS envelope width to very small values. At the same time the critical density was approaching zero. A similar effect has been later noticed studying the model mAbs molecules, confined in the hard-sphere fluid \cite{Hvozd2022a}.
In our calculation above, such a non-complete binding is modelled by variable $k$, varying the strength of the interaction. Experimentally, there are several ways to achieve this. The external parameters, which may be modified are: temperature, $p\mathrm{H}$, nature of the buffer and/or added electrolyte, as also the presence of other binding species (proteins).




\section{Conclusions}

In this paper we present a theoretical study of the model bi-specific antibodies forming a liquid. The data published in Refs.~\cite{Raut2015,Raut2016} provide useful guidance to behavior of the DVD Ig solutions, in particular with respect to the salt concentration, $p\mathrm{H}$, and nature of additives. Very valuable are data on the protein-protein interaction contained in the viscosity, the second virial coefficient, and 
the $T_{cloud}$ measurements.

As already mentioned above our current model is not designed to capture the influence of $p\mathrm{H}$, concentration and the nature of the added buffer, as also other subtle effects. Despite of these shortcomings  we have been able to qualitatively reproduce some important solution properties. In addition, we proposed an explanation for the very narrow width of the experimental liquid-liquid phase transition envelope. 

As a weakness of our  approach it might be considered the fact that we can model the viscosity behaviour with the particles having five fully bonding, while for the equally good agreement with the liquid-liquid phase separation data we need to make this interaction weaker ($k$ smaller than 1). At this point we shall stress the facts that viscosity is a dynamic while LLPS is a thermodynamic property. In particular, viscosity depends on the state of solvent, so it is strongly temperature dependent. The viscosity measurement were taken at $25^\circ$C, that is above the critical temperature.  Taking into addition into account also the other factors which may influence the liquid-liquid phase separation  (for example, the solution composition) the weakness mentioned above, is not that surprising.

\section{Acknowledgments}

TH and YVK acknowledge financial
support from the National Research Foundation of Ukraine
(Project no. 2020.02/0317) 
\bigskip
\bibliographystyle{achemso}

\bibliography{tadeja}

\providecommand{\latin}[1]{#1}
\makeatletter
\providecommand{\doi}
  {\begingroup\let\do\@makeother\dospecials
  \catcode`\{=1 \catcode`\}=2 \doi@aux}
\providecommand{\doi@aux}[1]{\endgroup\texttt{#1}}
\makeatother
\providecommand*\mcitethebibliography{\thebibliography}
\csname @ifundefined\endcsname{endmcitethebibliography}
  {\let\endmcitethebibliography\endthebibliography}{}
\begin{mcitethebibliography}{34}
\providecommand*\natexlab[1]{#1}
\providecommand*\mciteSetBstSublistMode[1]{}
\providecommand*\mciteSetBstMaxWidthForm[2]{}
\providecommand*\mciteBstWouldAddEndPuncttrue
  {\def\EndOfBibitem{\unskip.}}
\providecommand*\mciteBstWouldAddEndPunctfalse
  {\let\EndOfBibitem\relax}
\providecommand*\mciteSetBstMidEndSepPunct[3]{}
\providecommand*\mciteSetBstSublistLabelBeginEnd[3]{}
\providecommand*\EndOfBibitem{}
\mciteSetBstSublistMode{f}
\mciteSetBstMaxWidthForm{subitem}{(\alph{mcitesubitemcount})}
\mciteSetBstSublistLabelBeginEnd
  {\mcitemaxwidthsubitemform\space}
  {\relax}
  {\relax}

\bibitem[Thayer(2016)]{Thayer2016}
Thayer,~A.~M. Big pharma pursues next generation of antibodies. \emph{Chem. \&
  Eng. News} \textbf{2016}, \emph{94}, 14\relax
\mciteBstWouldAddEndPuncttrue
\mciteSetBstMidEndSepPunct{\mcitedefaultmidpunct}
{\mcitedefaultendpunct}{\mcitedefaultseppunct}\relax
\EndOfBibitem
\bibitem[Reichert(2017)]{Reichert2017}
Reichert,~J.~M. Antibodies to watch in 2017. \emph{mAbs} \textbf{2017},
  \emph{9}, 167--181\relax
\mciteBstWouldAddEndPuncttrue
\mciteSetBstMidEndSepPunct{\mcitedefaultmidpunct}
{\mcitedefaultendpunct}{\mcitedefaultseppunct}\relax
\EndOfBibitem
\bibitem[Starr and Tessier(2019)Starr, and Tessier]{Starr2019}
Starr,~C.; Tessier,~P.~M. Selecting and engineering monoclonal antibodies with
  drug-like specificity. \emph{Curr. Opin. Biotechnol.} \textbf{2019},
  \emph{60}, 119--127\relax
\mciteBstWouldAddEndPuncttrue
\mciteSetBstMidEndSepPunct{\mcitedefaultmidpunct}
{\mcitedefaultendpunct}{\mcitedefaultseppunct}\relax
\EndOfBibitem
\bibitem[Krieg \latin{et~al.}(2020)Krieg, Berner, Winter, and
  Svilenov]{Krieg2020}
Krieg,~D.; Berner,~C.; Winter,~G.; Svilenov,~H.~L. Biophysical Characterization
  of Binary Therapeutic Monoclonal Antibody Mixtures. \emph{Mol. Pharm.}
  \textbf{2020}, \emph{17}, 2971--2986\relax
\mciteBstWouldAddEndPuncttrue
\mciteSetBstMidEndSepPunct{\mcitedefaultmidpunct}
{\mcitedefaultendpunct}{\mcitedefaultseppunct}\relax
\EndOfBibitem
\bibitem[Le~Basle \latin{et~al.}(2020)Le~Basle, Chennell, Tokhadze, Astier, and
  Sautou]{Basle2020}
Le~Basle,~Y.; Chennell,~P.; Tokhadze,~N.; Astier,~A.; Sautou,~V.
  Physicochemical Stability of Monoclonal Antibodies: A Review. \emph{J. Pharm.
  Sci.} \textbf{2020}, \emph{109}, 169--190\relax
\mciteBstWouldAddEndPuncttrue
\mciteSetBstMidEndSepPunct{\mcitedefaultmidpunct}
{\mcitedefaultendpunct}{\mcitedefaultseppunct}\relax
\EndOfBibitem
\bibitem[Kontermann and Brinkmann(2015)Kontermann, and
  Brinkmann]{Kontermann2015}
Kontermann,~R.~E.; Brinkmann,~U. Bispecific antibodies. \emph{Drug Discov.
  Today} \textbf{2015}, \emph{20}, 838--847\relax
\mciteBstWouldAddEndPuncttrue
\mciteSetBstMidEndSepPunct{\mcitedefaultmidpunct}
{\mcitedefaultendpunct}{\mcitedefaultseppunct}\relax
\EndOfBibitem
\bibitem[Ma \latin{et~al.}(2021)Ma, Mo, Tang, Shen, Qi, Zhao, Huang, Hu, and
  Qian]{Ma2021}
Ma,~J.; Mo,~Y.; Tang,~M.; Shen,~J.; Qi,~Y.; Zhao,~W.; Huang,~Y.; Hu,~Y.;
  Qian,~C. Bispecific antibodies: from research to clinical application.
  \emph{Front. Immunol.} \textbf{2021}, \emph{899}, 626--616\relax
\mciteBstWouldAddEndPuncttrue
\mciteSetBstMidEndSepPunct{\mcitedefaultmidpunct}
{\mcitedefaultendpunct}{\mcitedefaultseppunct}\relax
\EndOfBibitem
\bibitem[S.~Gong \latin{et~al.}(2017)S.~Gong, Wu, Wu, and Wu]{Gong2017}
S.~Gong,~F.~R.; Wu,~D.; Wu,~X.; Wu,~C. Fabs-in-tandem immunoglobulin is a novel
  and versatile bispecific design for engaging multiple therapeutic targets.
  \emph{mAbs} \textbf{2017}, \emph{9}, 1118--1128\relax
\mciteBstWouldAddEndPuncttrue
\mciteSetBstMidEndSepPunct{\mcitedefaultmidpunct}
{\mcitedefaultendpunct}{\mcitedefaultseppunct}\relax
\EndOfBibitem
\bibitem[Gong and Wu(2019)Gong, and Wu]{Gong2019}
Gong,~S.; Wu,~C. Generation of Fabs-in-tandem immunoglobulin molecules for
  dual-specific targeting. \emph{Methods} \textbf{2019}, \emph{154},
  87--92\relax
\mciteBstWouldAddEndPuncttrue
\mciteSetBstMidEndSepPunct{\mcitedefaultmidpunct}
{\mcitedefaultendpunct}{\mcitedefaultseppunct}\relax
\EndOfBibitem
\bibitem[Svilenov \latin{et~al.}(2023)Svilenov, Arosio, Menzenc, Tessier, and
  Sormani]{Svilenova2023}
Svilenov,~H.; Arosio,~P.; Menzenc,~T.; Tessier,~P.; Sormani,~P. Approaches to
  expand the conventional toolbox for discovery and selection of antibodies
  with drug-like physicochemical properties. \emph{mAbs} \textbf{2023},
  \emph{15:}, 2164459\relax
\mciteBstWouldAddEndPuncttrue
\mciteSetBstMidEndSepPunct{\mcitedefaultmidpunct}
{\mcitedefaultendpunct}{\mcitedefaultseppunct}\relax
\EndOfBibitem
\bibitem[Chames and Baty(2009)Chames, and Baty]{Chames2009}
Chames,~P.; Baty,~D. Bispecific antibodies for cancer therapy: The light at the
  end of the tunnel. \emph{mAbs} \textbf{2009}, \emph{1}, 539--547\relax
\mciteBstWouldAddEndPuncttrue
\mciteSetBstMidEndSepPunct{\mcitedefaultmidpunct}
{\mcitedefaultendpunct}{\mcitedefaultseppunct}\relax
\EndOfBibitem
\bibitem[Correia \latin{et~al.}(2013)Correia, Sung, Burton, Jakob, Carragher,
  Ghayur, and Radziejewski]{Correia2013}
Correia,~I.; Sung,~J.; Burton,~R.; Jakob,~C.~G.; Carragher,~B.; Ghayur,~T.;
  Radziejewski,~C. The structure of dual--variable-domain immunoglobulin
  molecules (alone and bound to antigen). \emph{mAbs} \textbf{2013}, \emph{5},
  364--372\relax
\mciteBstWouldAddEndPuncttrue
\mciteSetBstMidEndSepPunct{\mcitedefaultmidpunct}
{\mcitedefaultendpunct}{\mcitedefaultseppunct}\relax
\EndOfBibitem
\bibitem[DiGiammarino \latin{et~al.}(2012)DiGiammarino, Ghayur, and
  Liu]{DiGiammarino2012}
DiGiammarino,~E.; Ghayur,~T.; Liu,~J. Design and generation of DVD-Ig™
  molecules for dual-specific targeting. \emph{Methods Mol Biol.}
  \textbf{2012}, \emph{899:}, 145--56\relax
\mciteBstWouldAddEndPuncttrue
\mciteSetBstMidEndSepPunct{\mcitedefaultmidpunct}
{\mcitedefaultendpunct}{\mcitedefaultseppunct}\relax
\EndOfBibitem
\bibitem[Jakob \latin{et~al.}(2013)Jakob, Edalji, Judge, DiGimmarino, Li, Gu,
  and Ghayur]{Jakob2013}
Jakob,~C.~G.; Edalji,~R.; Judge,~R.~A.; DiGimmarino,~E.; Li,~Y.; Gu,~J.;
  Ghayur,~T. Structure reveals function of the dual variable domain
  immunoglobulin (DVD-Ig™) molecule. \emph{mAbs} \textbf{2013}, \emph{5},
  358--363\relax
\mciteBstWouldAddEndPuncttrue
\mciteSetBstMidEndSepPunct{\mcitedefaultmidpunct}
{\mcitedefaultendpunct}{\mcitedefaultseppunct}\relax
\EndOfBibitem
\bibitem[Raut and Kalonia(2015)Raut, and Kalonia]{Raut2015}
Raut,~A.~S.; Kalonia,~D.~S. Liquid--Liquid Phase Separation in a Dual Variable
  Domain Immunoglobulin Protein Solution: Effect of Formulation Factors and
  Protein--Protein Interactions. \emph{Mol. Pharm.} \textbf{2015}, \emph{12},
  3261--3271\relax
\mciteBstWouldAddEndPuncttrue
\mciteSetBstMidEndSepPunct{\mcitedefaultmidpunct}
{\mcitedefaultendpunct}{\mcitedefaultseppunct}\relax
\EndOfBibitem
\bibitem[Raut and Kalonia(2016)Raut, and Kalonia]{Raut2016}
Raut,~A.~S.; Kalonia,~D.~S. Effect of Excipients on Liquid--Liquid Phase
  Separation and Aggregation in Dual Variable Domain Immunoglobulin Protein
  Solutions. \emph{Mol. Pharm.} \textbf{2016}, \emph{13}, 774--783\relax
\mciteBstWouldAddEndPuncttrue
\mciteSetBstMidEndSepPunct{\mcitedefaultmidpunct}
{\mcitedefaultendpunct}{\mcitedefaultseppunct}\relax
\EndOfBibitem
\bibitem[Raut and Kalonia(2016)Raut, and Kalonia]{Raut2016a}
Raut,~A.~S.; Kalonia,~D.~S. Viscosity Analysis of Dual Variable Domain
  Immunoglobulin Protein Solutions: Role of Size, Electroviscous Effect and
  Protein-Protein Interactions. \emph{Pharm. Res.} \textbf{2016}, \emph{35},
  155--166\relax
\mciteBstWouldAddEndPuncttrue
\mciteSetBstMidEndSepPunct{\mcitedefaultmidpunct}
{\mcitedefaultendpunct}{\mcitedefaultseppunct}\relax
\EndOfBibitem
\bibitem[Butovych \latin{et~al.}(2023)Butovych, Kalyuzhnyi, Patsahan, and
  Ilnytskyi]{Butovych2023}
Butovych,~H.; Kalyuzhnyi,~Y.~V.; Patsahan,~T.; Ilnytskyi,~J. Modeling of
  polymer-enzyme conjugates formation: Thermodynamic perturbation theory and
  computer simulations. \emph{J. Mol. Liq.} \textbf{2023}, 122321\relax
\mciteBstWouldAddEndPuncttrue
\mciteSetBstMidEndSepPunct{\mcitedefaultmidpunct}
{\mcitedefaultendpunct}{\mcitedefaultseppunct}\relax
\EndOfBibitem
\bibitem[Wertheim(1986)]{Wertheim3}
Wertheim,~M.~S. Fluids with highly directional attractive forces III. Multiple
  attraction sites. \emph{J. Stat. Phys.} \textbf{1986}, \emph{42},
  459--476\relax
\mciteBstWouldAddEndPuncttrue
\mciteSetBstMidEndSepPunct{\mcitedefaultmidpunct}
{\mcitedefaultendpunct}{\mcitedefaultseppunct}\relax
\EndOfBibitem
\bibitem[Wertheim(1986)]{Wertheim4}
Wertheim,~M.~S. Fluids with highly directional attractive forces. {IV}.
  {E}quilibrium polymerization. \emph{J. Stat. Phys.} \textbf{1986}, \emph{42},
  477--492\relax
\mciteBstWouldAddEndPuncttrue
\mciteSetBstMidEndSepPunct{\mcitedefaultmidpunct}
{\mcitedefaultendpunct}{\mcitedefaultseppunct}\relax
\EndOfBibitem
\bibitem[Wertheim(1987)]{Wertheim1987}
Wertheim,~M.~S. Thermodynamic perturbation theory of polymerization. \emph{J.
  Chem. Phys.} \textbf{1987}, \emph{87}, 7323--7331\relax
\mciteBstWouldAddEndPuncttrue
\mciteSetBstMidEndSepPunct{\mcitedefaultmidpunct}
{\mcitedefaultendpunct}{\mcitedefaultseppunct}\relax
\EndOfBibitem
\bibitem[Kastelic \latin{et~al.}(2016)Kastelic, Kalyuzhnyi, and
  Vlachy]{Kastelic2016b}
Kastelic,~M.; Kalyuzhnyi,~Y.~V.; Vlachy,~V. Modeling phase transitions in
  mixtures of $\beta$--$\gamma$ lens crystallins. \emph{Soft Matter}
  \textbf{2016}, \emph{12}, 7289--7298\relax
\mciteBstWouldAddEndPuncttrue
\mciteSetBstMidEndSepPunct{\mcitedefaultmidpunct}
{\mcitedefaultendpunct}{\mcitedefaultseppunct}\relax
\EndOfBibitem
\bibitem[Kastelic \latin{et~al.}(2017)Kastelic, Kalyuzhnyi, Dill, and
  Vlachy]{Kastelic2017}
Kastelic,~M.; Kalyuzhnyi,~Y.; Dill,~K.~A.; Vlachy,~V. Controlling the
  viscosities of antibody solutions through control of their binding sites.
  \emph{J. Mol. Liq.} \textbf{2017}, \emph{270}, 234--242\relax
\mciteBstWouldAddEndPuncttrue
\mciteSetBstMidEndSepPunct{\mcitedefaultmidpunct}
{\mcitedefaultendpunct}{\mcitedefaultseppunct}\relax
\EndOfBibitem
\bibitem[Kastelic and Vlachy(2018)Kastelic, and Vlachy]{Kastelic2018}
Kastelic,~M.; Vlachy,~V. Theory for the Liquid--Liquid Phase Separation in
  Aqueous Antibody Solutions. \emph{J. Phys. Chem. B} \textbf{2018},
  \emph{122}, 5400--5408\relax
\mciteBstWouldAddEndPuncttrue
\mciteSetBstMidEndSepPunct{\mcitedefaultmidpunct}
{\mcitedefaultendpunct}{\mcitedefaultseppunct}\relax
\EndOfBibitem
\bibitem[Hvozd \latin{et~al.}(2020)Hvozd, Kalyuzhnyi, and Vlachy]{Hvozd2020}
Hvozd,~T.; Kalyuzhnyi,~Y.~V.; Vlachy,~V. Aggregation, liquid--liquid phase
  separation, and percolation behaviour of a model antibody fluid constrained
  by hard-sphere obstacles. \emph{Soft Matter} \textbf{2020}, \emph{16},
  8432--8443\relax
\mciteBstWouldAddEndPuncttrue
\mciteSetBstMidEndSepPunct{\mcitedefaultmidpunct}
{\mcitedefaultendpunct}{\mcitedefaultseppunct}\relax
\EndOfBibitem
\bibitem[Hvozd \latin{et~al.}(2022)Hvozd, Kalyuzhnyi, Vlachy, and
  Cummings]{Hvozd2022}
Hvozd,~T.~V.; Kalyuzhnyi,~Y.~V.; Vlachy,~V.; Cummings,~P.~T. Empty liquid state
  and re-entrant phase behavior of the patchy colloids confined in porous
  medias. \emph{J. Chem. Phys.} \textbf{2022}, \emph{156}, 1--5\relax
\mciteBstWouldAddEndPuncttrue
\mciteSetBstMidEndSepPunct{\mcitedefaultmidpunct}
{\mcitedefaultendpunct}{\mcitedefaultseppunct}\relax
\EndOfBibitem
\bibitem[Kalyuzhnyi and Cummings(1995)Kalyuzhnyi, and Cummings]{Kalyuzhnyi1995}
Kalyuzhnyi,~Y.~V.; Cummings,~P. Solution of the polymer Percus--Yevick
  approximation for the multicomponent totally flexible sticky two-point model
  of polymerizing fluid. \emph{J. Chem. Phys.} \textbf{1995}, \emph{103},
  3265--3267\relax
\mciteBstWouldAddEndPuncttrue
\mciteSetBstMidEndSepPunct{\mcitedefaultmidpunct}
{\mcitedefaultendpunct}{\mcitedefaultseppunct}\relax
\EndOfBibitem
\bibitem[Kalyuzhnyi \latin{et~al.}(1997)Kalyuzhnyi, Lin, and
  Stell]{Kalyuzhnyi1997}
Kalyuzhnyi,~Y.~V.; Lin,~C.-T.; Stell,~G. Primitive models of chemical
  association. II. Polymerization into flexible chain molecules of prescribed
  length. \emph{J. Chem. Phys.} \textbf{1997}, \emph{106}, 1940--1949\relax
\mciteBstWouldAddEndPuncttrue
\mciteSetBstMidEndSepPunct{\mcitedefaultmidpunct}
{\mcitedefaultendpunct}{\mcitedefaultseppunct}\relax
\EndOfBibitem
\bibitem[Lin \latin{et~al.}(1998)Lin, Kalyuzhnyi, and Stell]{Lin1998}
Lin,~C.-T.; Kalyuzhnyi,~Y.~V.; Stell,~G. Primitive models of chemical
  association. III. Totally flexible sticky two-point model for multicomponent
  heteronuclear fixed-chain-length polymerization. \emph{J. Chem. Phys.}
  \textbf{1998}, \emph{108}, 6513--6524\relax
\mciteBstWouldAddEndPuncttrue
\mciteSetBstMidEndSepPunct{\mcitedefaultmidpunct}
{\mcitedefaultendpunct}{\mcitedefaultseppunct}\relax
\EndOfBibitem
\bibitem[Bianchi \latin{et~al.}(2006)Bianchi, Lartgo, Tartaglia, Zaccarelli,
  and Sciortino]{Bianchi2006}
Bianchi,~E.; Lartgo,~J.; Tartaglia,~P.; Zaccarelli,~E.; Sciortino,~F. Phase
  diagram of patchy colloids: Towards empty liquids. \emph{Phys. Rev. Lett.}
  \textbf{2006}, \emph{97}, 168301\relax
\mciteBstWouldAddEndPuncttrue
\mciteSetBstMidEndSepPunct{\mcitedefaultmidpunct}
{\mcitedefaultendpunct}{\mcitedefaultseppunct}\relax
\EndOfBibitem
\bibitem[George and Wilson(1994)George, and Wilson]{George1994}
George,~A.; Wilson,~W.~W. Predicting protein crystallization from a
  dilute--solution property. \emph{Acta Crystallogr. D} \textbf{1994},
  \emph{50}, 361--365\relax
\mciteBstWouldAddEndPuncttrue
\mciteSetBstMidEndSepPunct{\mcitedefaultmidpunct}
{\mcitedefaultendpunct}{\mcitedefaultseppunct}\relax
\EndOfBibitem
\bibitem[Tomar \latin{et~al.}(2018)Tomar, Singh, Li~Li, and Kumar]{Tomar2018}
Tomar,~D.~S.; Singh,~S.~K.; Li~Li,~M. P.~B.; Kumar,~S. In Silico Prediction of
  Diffusion Interaction Parameter ($k_D$), a Key Indicator of Antibody Solution
  Behaviors. \emph{Pharm. Res.} \textbf{2018}, \emph{35}, 193, 1--20\relax
\mciteBstWouldAddEndPuncttrue
\mciteSetBstMidEndSepPunct{\mcitedefaultmidpunct}
{\mcitedefaultendpunct}{\mcitedefaultseppunct}\relax
\EndOfBibitem
\bibitem[Kalyuzhnyi and Vlachy(2016)Kalyuzhnyi, and Vlachy]{Kalyuzhnyi2016}
Kalyuzhnyi,~Y.~V.; Vlachy,~V. Explicit--water theory for the salt--specific
  effects and {H}ofmeister series in protein solutions. \emph{J. Chem. Phys.}
  \textbf{2016}, \emph{144}, 215101\relax
\mciteBstWouldAddEndPuncttrue
\mciteSetBstMidEndSepPunct{\mcitedefaultmidpunct}
{\mcitedefaultendpunct}{\mcitedefaultseppunct}\relax
\EndOfBibitem
\bibitem[Hvozd \latin{et~al.}(2022)Hvozd, Kalyuzhnyi, and Vlachy]{Hvozd2022a}
Hvozd,~T.; Kalyuzhnyi,~Y.~V.; Vlachy,~V. Behaviour of the model antibody fluid
  constrained by rigid spherical obstacles : effects of the obstacle–antibody
  attraction. \emph{Soft Matter} \textbf{2022}, \emph{18}, 9108--9117\relax
\mciteBstWouldAddEndPuncttrue
\mciteSetBstMidEndSepPunct{\mcitedefaultmidpunct}
{\mcitedefaultendpunct}{\mcitedefaultseppunct}\relax
\EndOfBibitem
\end{mcitethebibliography}


\begin{thebibliography}{11}
\expandafter\ifx\csname natexlab\endcsname\relax\def\natexlab#1{#1}\fi
\providecommand{\url}[1]{\texttt{#1}}
\providecommand{\href}[2]{#2}
\providecommand{\path}[1]{#1}
\providecommand{\DOIprefix}{doi:}
\providecommand{\ArXivprefix}{arXiv:}
\providecommand{\URLprefix}{URL: }
\providecommand{\Pubmedprefix}{pmid:}
\providecommand{\doi}[1]{\href{http://dx.doi.org/#1}{\path{#1}}}
\providecommand{\Pubmed}[1]{\href{pmid:#1}{\path{#1}}}
\providecommand{\bibinfo}[2]{#2}
\ifx\xfnm\relax \def\xfnm[#1]{\unskip,\space#1}\fi
\bibitem[{Collins(1997)}]{Collins1997}
\bibinfo{author}{K.~D. Collins}, \bibinfo{journal}{Biophysical Journal}
  \bibinfo{volume}{72} (\bibinfo{year}{1997}) \bibinfo{pages}{65--76}.
\bibitem[{Lyklema(2003)}]{Lyklema2003}
\bibinfo{author}{J.~Lyklema}, \bibinfo{journal}{Advances in Colloid and
  Interface Science} \bibinfo{volume}{100-102} (\bibinfo{year}{2003})
  \bibinfo{pages}{1--12}.
\bibitem[{Kunz and Neueder(2010)}]{Kunz2009}
\bibinfo{author}{W.~Kunz}, \bibinfo{author}{R.~Neueder},
  \bibinfo{title}{Specific Ion Effects}, \bibinfo{publisher}{World Scientific,
  Singapore}, \bibinfo{year}{2010}, pp. \bibinfo{pages}{3--54}.
\bibitem[{Kunz(2010)}]{Kunz2010}
\bibinfo{author}{W.~Kunz}, \bibinfo{journal}{Current Opinion in Colloid \&
  Interface Science} \bibinfo{volume}{15} (\bibinfo{year}{2010})
  \bibinfo{pages}{34--39}.
\bibitem[{Lo~Nostro and Ninham(2012)}]{lonostro}
\bibinfo{author}{P.~Lo~Nostro}, \bibinfo{author}{B.~W. Ninham},
  \bibinfo{journal}{Chemical Reviews} \bibinfo{volume}{112}
  (\bibinfo{year}{2012}) \bibinfo{pages}{2286--2322}.
\bibitem[{Lah et~al.(2001)Lah, Maier, Lindner, and Vesnaver}]{Lah3}
\bibinfo{author}{J.~Lah}, \bibinfo{author}{N.~M. Maier},
  \bibinfo{author}{W.~Lindner}, \bibinfo{author}{G.~Vesnaver},
  \bibinfo{journal}{The Journal of Physical Chemistry B} \bibinfo{volume}{105}
  (\bibinfo{year}{2001}) \bibinfo{pages}{1670--1678}.
\bibitem[{Bončina et~al.(2010)Bončina, Lah, Reščič, and Vlachy}]{Matjaz}
\bibinfo{author}{M.~Bončina}, \bibinfo{author}{J.~Lah},
  \bibinfo{author}{J.~Reščič}, \bibinfo{author}{V.~Vlachy},
  \bibinfo{journal}{The Journal of Physical Chemistry B} \bibinfo{volume}{114}
  (\bibinfo{year}{2010}) \bibinfo{pages}{4313--4319}.
\bibitem[{Lah et~al.(2003)Lah, Lah, Zegers, Wyns, and Messens}]{LahN}
\bibinfo{author}{N.~Lah}, \bibinfo{author}{J.~Lah},
  \bibinfo{author}{I.~Zegers}, \bibinfo{author}{L.~Wyns},
  \bibinfo{author}{J.~Messens}, \bibinfo{journal}{The Journal of Biological
  Chemistry} \bibinfo{volume}{278} (\bibinfo{year}{2003})
  \bibinfo{pages}{24673--24679}.
\bibitem[{Lah et~al.(2006)Lah, Carl, Drobnak, Šumiga, and Vesnaver}]{Lah1}
\bibinfo{author}{J.~Lah}, \bibinfo{author}{N.~Carl},
  \bibinfo{author}{I.~Drobnak}, \bibinfo{author}{B.~Šumiga},
  \bibinfo{author}{G.~Vesnaver}, \bibinfo{journal}{Acta Chimica Slovenica}
  \bibinfo{volume}{53} (\bibinfo{year}{2006}) \bibinfo{pages}{284--291}.
\bibitem[{Lah and Vesnaver(2004)}]{Lah2}
\bibinfo{author}{J.~Lah}, \bibinfo{author}{G.~Vesnaver},
  \bibinfo{journal}{Journal of Molecular Biology} \bibinfo{volume}{342}
  (\bibinfo{year}{2004}) \bibinfo{pages}{73--89}.
\bibitem[{Lah et~al.(2006)Lah, Drobnak, Dolinar, and Vesnaver}]{Lah4}
\bibinfo{author}{J.~Lah}, \bibinfo{author}{I.~Drobnak},
  \bibinfo{author}{M.~Dolinar}, \bibinfo{author}{G.~Vesnaver},
  \bibinfo{journal}{Nucleic Acid Research} \bibinfo{volume}{36}
  (\bibinfo{year}{2006}) \bibinfo{pages}{897--904}.

\end{thebibliography}
\end{document}